\documentclass[journal, 10pt]{IEEEtran}

\usepackage{cite}
\usepackage{mdwtab}
\usepackage{graphicx}
\usepackage{amsmath}
\usepackage{mathtools}
\usepackage{mathdots}
\usepackage{amsfonts}
\usepackage{latexsym}
\usepackage{subfigure}
\usepackage{placeins}
\usepackage{psfrag}
\usepackage{setspace}
\usepackage{stmaryrd}
\usepackage{epsfig}
\usepackage{epstopdf}
\usepackage[latin1]{inputenc}
\usepackage{amssymb}
\usepackage{amsthm}
\usepackage{url}
\usepackage{lipsum}
\usepackage{multicol}
\usepackage{tikz}
\usepackage{cancel}
\usepackage{verbatim}

\allowdisplaybreaks

\theoremstyle{plain}
\theoremstyle{remark}

\begin{document}
\newtheorem{theorem}{Theorem}[section]
\newtheorem{lemma}{Lemma}
\newtheorem{conjecture}{Conjecture}
\newtheorem{corollary}{Corollary}
\newtheorem{definition}{Definition}
\newtheorem{property}{Property}

\bibliographystyle{plain}

\title{A New NOMA Approach for Fair Power Allocation }

\author{\IEEEauthorblockN{Jos\'{e} Armando Oviedo and  Hamid R. Sadjadpour}\\
\IEEEauthorblockA{Department of Electrical Engineering, University of California, Santa Cruz\\
Email: \{xmando, hamid\}@soe.ucsc.edu}}

\maketitle

\IEEEpeerreviewmaketitle

\begin{abstract}
A non-orthogonal multiple access (NOMA) approach to user signal power allocation called Fair-NOMA is introduced. Fair-NOMA is the application of NOMA in such a way that two mobile users have the opportunity to always achieve at least the information capacity they can achieve by using orthogonal multiple access (OMA), regardless of the user selection criteria, making it suitable for implementation using any current or future  scheduling paradigms. Given this condition, the bounds of the power allocation coefficients are derived as functions of the channel  gains of the two mobile users. The NOMA power allocation is  analyzed for {\it two scheduled users} that are selected randomly with i.i.d. channel gains. The capacity improvements made by each user and the sum capacity improvement are  derived.

\end{abstract}

\section{Introduction}

A system that employs \textit{orthogonal multiple access} (OMA) is defined as a system that schedules multiple mobile users (MUs) in non-overlapping timeslots or frequency bands during a certain transmission time period. Therefore, if the signals for users MU-$k$, $k=1,\ldots K$ are scheduled to be transmitted over a time period $T$, where $T$ is less than the coherence time of the channel, then both MU-1 and MU-2 have their signals transmitted only $T/K$ amount of the total  transmission period (or fraction of the total bandwidth). Since only one of the signals is transmitted at any given time slot (or frequency band), that particular signal is allocated all of the transmit SNR $\xi$.

A system that employs \textit{non-orthogonal multiple access} (NOMA) is one that, given the same users as above, schedules the transmission of their signals over the entire transmission period and bandwidth by using superposition coding (SC). However, since the total transmit SNR $\xi $ must be shared between the $k$ signals being transmitted, a fraction $a_k\in(0,1)$ of the transmit power is allocated to user $k$, and $\sum_{k=1}^K a_k = 1$. In order for NOMA to be viable approach to scheduling users, each user must employ successive interference cancellation (SIC) at the receiver to remove the interference of the signals from users that have lesser channel SNR gains \cite{InfTh:CT}.

An approach called Fair-NOMA is proposed for two users for future wireless cellular downlink systems as a framework to implement NOMA fairly. The underlying fundamental property of Fair-NOMA is that  users will always be guaranteed to achieve a capacity at least as good as OMA. This is achieved by deriving the exact bounds for the Fair-NOMA power allocation region $\mathcal{A}_\text{FN} = [a_\text{inf}, a_\text{sup}]\subseteq [0,1]$, as functions of the channel gains of the scheduled MUs. For the case where two random users with i.i.d. channel SNR gains are selected, the average capacities for both weaker and stronger users are derived at each lower and upper bound on the Fair-NOMA power allocation region. The expected increase in capacity between OMA and NOMA are derived for each bound as well, which provides insight as to how  much the capacity improves even with the restriction imposed by $\mathcal{A}_\text{FN}$.

The practicality of the Fair-NOMA approach is that  it requires the receivers to possess the ability to perform SIC. It must be stressed that Fair-NOMA will always improve the sum capacity of the network \textit{and} the capacity of each individual user compared to OMA. Furthermore, Fair-NOMA does not require any additional feedback when compared to other NOMA techniques, which is  the absolute value of the gain (no channel phase information is required). Therefore, there is no need to discuss the probability of NOMA failing to improve capacity performance, and we can focus only on how much capacity gain will provide. A simple analysis of outage capacity is briefly discussed to provide a more thorough treatment of the performance of Fair-NOMA.

Another unique feature of our approach compared to the previous work is the fact that prior studies on NOMA have  focused on demonstrating that NOMA has advantages for increasing the capacity of the network when users are scheduled and paired based on their channel conditions (i.e. their location in the cell). Fair-NOMA does not rely on this condition in its analysis and simulation, since users' channel conditions are i.i.d. distributed (i.e. location in the cell is not considered). Hence, all users will have equal opportunity to be scheduled, and thus is also completely "fair" from a time-sharing perspective. However, Fair-NOMA can be applied to any system with any scheduling and user-pairing approach.

The paper is organized as follows. The discussion of the important previous work on the development of the NOMA concept is outlined in section \ref{sec2}. The  system model is outlined in section \ref{sec3}. Section \ref{sec4} defines the Fair-NOMA power allocation region $\mathcal{A}_\text{FN}$, and develops its basic properties. The analysis of the effects of Fair-NOMA on the capacity of each user is provided in section \ref{sec5} for the boundary power allocation coefficient values, and simulation results verify the analysis and demonstrate the performance improvement. Finally, section \ref{sec7} concludes the paper and discusses the future work to be considered.

\section{Previous Work on NOMA} \label{sec2}

The concept of NOMA is based on using superposition coding (SC) at the transmitter and successive interference cancellation (SIC) at the receivers. This was shown to achieve the capacity of the channel by Cover and Thomas \cite{InfTh:CT}. The existence of a set of power allocation coefficients that allow all of the participating users to achieve capacity at least as good as OMA was suggested in \cite{FundWiCom:Tse}. 
With advances in computing technology, it is reasonable to suggest that a mobile receiver will possess the capability to perform the required SIC operation, making NOMA an attractive option for implementation in future wireless standards \cite{Li-2014}.

Non-orthogonal access approaches using SC for future wireless cellular networks was mentioned in \cite{CompOMANOMA:WXP} as a way to increase single user rates when compared to CDMA. Schaepperle and Ruegg \cite{4GNOSig:SR} evaluated the performance of non-orthogonal signaling using SC and SIC in single antenna OFDMA systems using very little modifications to the existing standards, as well as how user pairing impacts the throughput of the system when the channel gains become increasingly disparate. This was then applied by Schaepperle \cite{WCSCMA:Schaep} to OFDMA wireless systems to evaluate the performance of cell edge user rates, proposing an algorithm that attempts to increase the average throughput and maintain fairness. These works do not assume to have the exact channel state information at the transmitter.

The concept of NOMA is  evaluated through simulation for full CSIT in the uplink \cite{ULNOMA:TakedaHiguchi} and downlink \cite{DLNOMA:TomidaHiguchi}, where the throughput of the system is shown to be on average always better for NOMA than OMA when considering a fully defined cellular system evaluation, with both users occupying all of the bandwidth and time, and was compared to FDMA with each user being assigned an orthogonal channel. In \cite{SLDLNOMA:Saitoetal}, the downlink system performance throughput gains are evaluated by incorporating a complete simulation of an LTE cellular system (3GPP). Further simulation studies were done to evaluate the performance of NOMA for scheduling multiple users per sub-band in OFDMA systems \cite{FDULNOM:Gotoetal}, and it is shown that when scheduling users, the users selected in each sub-band are determined by predicting which sub-band each user should be in, such that the expected throughput is maximized.

{
Fairness in NOMA systems is addressed in some works. The uplink case in OFDMA systems is addressed in \cite{5GNOMAUp:AXIT} by using an algorithm that attempts to maximize the sum throughput, with respect to OFDMA and power constraints. The fairness is not directly addressed in the problem formulation, but is evaluated using Jain's fairness index. In   \cite{PropFairNOMA:LMP}, a proportional fair scheduler and user pair power allocation scheme is used to achieve fairness in time and rate.  In \cite{FairnessNOMA5G:TK}, fairness is achieved in the max-min sense, where users are paired such that their channel conditions are not too disparate, while the power allocation maximizes the rates for the paired users. A closed-form solution is reached for the instantaneous CSIT case, and an efficient algorithm is found for the case with average CSIT. }

Ping et. al. \cite{5GNOMA:DFP} provide an analysis for fixed-power NOMA, where the power allocation coefficient is fixed for the weaker "cell-edge" user at $a_m=4/5$ and for the stronger "near" user at $a_n=1/5$, and it is shown that the probability that NOMA outperforms OMA approaches $1$ as the number of users in the network increases.

{Our main contribution is to define the exact power allocation region that will allow for implementation of NOMA to any system in a "fair" manner. We define "fair" here as being a technique where all scheduled users have a capacity equal or greater than OMA. In other words, no proportional fair schedulers are required to guarantee per-user capacity is always at least better than the OMA case, while {\em user selection bias} is taken from channel conditions or previous rates. Like many NOMA techniques, we require full CSIT in order for the BS to properly perform the superposition coding, while the users only need to be notified of the rates and modulation used for each signal to enable SIC (if needed). It is important to note that our NOMA approach {\em always} guarantees equal or higher capacity than OMA. 
}

%%%%%%%%%%%%%%%%%%%%%%%%%%%%%%%%%%%%%%%%%%%%%%%%%%%%
%%%%%%%%%%%%%%%%%%%%%%%%%%%%%%%%%%%%%%%%%%%%%%%%%%%%
%%%%%%%%%%%%%%%%%%%%%%%%%%%%%%%%%%%%%%%%%%%%%%%%%%%%
%%
%%%%%%%%%%%%%%	  Fair-NOMA		%%%%%%%%%%%%%%%%%%%
%%
%%%%%%%%%%%%%%%%%%%%%%%%%%%%%%%%%%%%%%%%%%%%%%%%%%%%
%%%%%%%%%%%%%%%%%%%%%%%%%%%%%%%%%%%%%%%%%%%%%%%%%%%%
%%%%%%%%%%%%%%%%%%%%%%%%%%%%%%%%%%%%%%%%%%%%%%%%%%%%

\section{System Model } \label{sec3}

Let a mobile user MU-$i$ have a signal $x_i$ transmitted from a single antenna base-station (BS). The channel gain is $h_i\in\mathbb{C}$ with SNR gain p.d.f. $f_{|h|^2}(w) = \frac{1}{\beta}e^{-\frac{w}{\beta}}$, and receiver noise $z_i\sim\mathcal{CN}(0,1)$. If MU-1 and MU-2 each have their signals transmitted, with total transmit SNR $\xi$, each during half of the time period $T$ using OMA scheduling, then the received signal for each user in their respective half of the time period is $y_i = h_i \xi  x_i + z_i, i=1,2$. If $\mathbb{E}[ |x_i|^2 ] = 1$, the information capacity of each user is then 
\begin{equation}
	C_i^\text{O} = \frac{1}{2}\log_2\left(1 + \xi |h_i|^2\right), 
\end{equation}
where the $\frac{1}{2}$ factor accounts for the fact that each user has the available channel only half the time. In the case of NOMA, where both signals are being transmitted simultaneously during the entire time period $T$, the user with greater channel gain, which we assume to be MU-2 w.l.o.g., can perform SIC at the receiver by first treating its own signal as noise and decoding MU-1's signal. If the power allocation coefficient for MU-2 is $a\in(0,1/2)$, then MU-1's signal is allocated $1-a$ transmit power, and the received signals for both users are
\begin{align}
	y_1 &= \sqrt{(1-a)\xi }h_1 x_1 + \sqrt{a\xi }h_1 x_2 + z_1 \nonumber\\
	y_2 &= \sqrt{a\xi }h_2 x_2 + \sqrt{(1-a)\xi }h_2 x_1 + z_2 .
\end{align}
Since $|h_2|^2>|h_1|^2$, then
\begin{equation}
	\dfrac{a\xi |h_2|^2}{(1-a)\xi |h_2|^2+1} > \dfrac{a\xi |h_1|^2}{(1-a)\xi |h_1|^2+1},
\end{equation}
MU-2's receiver will perform SIC and remove the interference from MU-1's signal. Doing so, the capacity for each user in NOMA is
\begin{align}
		&C_1^\text{N}(a) = \log_2\left( 1 + \dfrac{(1-a)\xi |h_1|^2}{a\xi |h_1|^2 + 1}\right)\\
		&C_2^\text{N}(a) = \log_2\left( 1 + a\xi |h_2|^2\right).
\end{align}

By directly comparing the capacities such that we want $C_1^\text{N}(a)\geq C_1^\text{O}$ and  $C_2^\text{N}(a)\geq C_2^\text{O}$, the region that contains the values of $a$ can be easily found.

\section{Fair-NOMA Power Allocation Region} \label{sec4}

For MU-1, the power allocation coefficient $a$ that ensures that $C_1^\text{N}(a)\geq C_1^\text{O}$ is found by solving
\begin{equation}
	\log_2\left( 1 + \dfrac{(1-a)\xi |h_1|^2}{a\xi |h_1|^2 + 1}\right) \geq \frac{1}{2}\log_2\left(1 + \xi |h_1|^2\right)
\end{equation}
%Comparing the arguments of the logarithms directly results in
%\begin{equation}
%	1+\dfrac{(1-a)\xi |h_1|^2}{a\xi |h_1|^2 + 1} \geq \left(1 + \xi |h_1|^2\right)^\frac{1}{2}.
%\end{equation}
Solving the above inequality for $a$ gives 
\begin{align}
	& \Rightarrow \label{eq:asup}  a \leq \dfrac{(1 + \xi |h_1|^2)^{1/2} - 1}{\xi |h_1|^2} .
\end{align}
Therefore, the greatest value of the power allocation coefficient $a$ to ensure that NOMA is fair to MU-1 is given by the right side of (\ref{eq:asup}), and any $a$ satisfying (\ref{eq:asup}) will lead to $C_1^\text{N}(a)\geq C_1^\text{O}$.

Similarly, if the capacity of MU-2 using NOMA is to be at least as good as OMA, then $C_2^\text{N}(a)\geq C_2^\text{O}$ leads to
\begin{align}
%	&1 + a\xi |h_2|^2 \geq \left(1 + \xi |h_2|^2\right)^{1/2}\nonumber\\
	\label{eq:ainf} a \geq \dfrac{(1 + \xi |h_2|^2)^{1/2} - 1}{\xi |h_2|^2}.
\end{align}
Therefore, the least value of power allocation coefficient $a$ such that $C_2^\text{N}(a)\geq C_2^\text{O}$ is given by the right side of (\ref{eq:ainf}).

Each of the above values of $a$ that ensure fairness in capacity performance have the form of the function $a(x) = [(1 + \xi x)^{1/2}-1]/(\xi x)$. 

\begin{property}\label{prop:ainfsup}
	For a channel SNR gain $x$, the function $a(x)$ is a monotonically decreasing function of $x$, and $a(x)\in(0,1/2)$. 
\end{property}
\begin{IEEEproof}
	If $a(x)$ is a monotonically decreasing function of $x$, where $x>0$, then we must have $ \frac{da(x)}{dx} < 0$. 
	\begin{equation}
		\frac{da(x)}{dx} = -\dfrac{\frac{1}{2}\xi x + 1 - (1+\xi x)^{1/2} }{\xi x^2(1+\xi x)^{1/2}}
	\end{equation}
	It is easy to show that both  the numerator and denominator are positive $\forall x>0$,  proving that $\frac{da(x)}{dx} < 0$.
	To prove that $a(x)\in(0, 1/2)$, we have $\lim_{x\rightarrow b}a(x) = \frac{1}{2}(1 + \xi x)^{-1/2}$. 
%\begin{align}\label{eq:a_limit}
%	 & \lim_{x\rightarrow a}\dfrac{(1 + \xi x)^{1/2} - 1}{\xi x} \nonumber\\
%	\stackrel{*}{=}& \lim_{x\rightarrow a}\dfrac{\frac{1}{2}(1 + \xi x)^{-1/2}(\xi)}{\xi }\nonumber\\
%	=& \lim_{x\rightarrow a}\dfrac{\frac{1}{2}}{(1 + \xi x)^{1/2}},
%\end{align}
%where $*$ is true by l'H\^{o}pital's rule. 
Taking the limit as $x\rightarrow 0$ gives 
\begin{equation}
	\lim_{x\rightarrow 0}\dfrac{1}{2(1 + \xi x)^{1/2}} = \frac{1}{2},
\end{equation}
while taking the limit as $x\rightarrow \infty$ gives
\begin{equation}
	\lim_{x\rightarrow \infty}\dfrac{1}{2(1 + \xi x)^{1/2}} = 0.
\end{equation}
Hence, $a(x)$ is a monotonically decreasing function of $x$ in the range $(0,1/2)$. 
\end{IEEEproof}

Define $a_{\inf} = [(1 + \xi |h_2|^2)^{1/2} - 1]/ (\xi |h_2|^2)$ and $a_{\sup} = [(1 + \xi |h_1|^2)^{1/2} - 1]/(\xi |h_1|^2)$. Then by Property \ref{prop:ainfsup}, it is clear that if $|h_1|^2<|h_2|^2 \Rightarrow a_\text{inf} < a_\text{sup}$. The Fair-NOMA power allocation region is therefore defined as $\mathcal{A}_\text{FN}=[a_\text{inf}, a_\text{sup}]$, and selecting any $a\in\mathcal{A}_\text{FN}$ gives
\begin{align}
	& C_1^\text{N}(a)\geq C_1^\text{O},\nonumber\\
	& C_2^\text{N}(a)\geq C_2^\text{O},\nonumber\\
	\label{eq:FairNOMA}& S_\text{N}(a) > S_\text{O}.
\end{align}
Since the sum capacity $S_\text{N}(a) = C_1^\text{N}(a)+C_2^\text{N}(a)$ is a monotonically increasing function of $a$, then $a_\text{sup} = \arg{\displaystyle\max_{a\in\mathcal{A}_\text{FN}}}(C_2^\text{N}(a))$ also maximizes $S_\text{N}(a)$ when $a\in\mathcal{A}_\text{FN}$. The last inequality is strict because since at the least one of the MU's capacities always increases, then the sum capacity always increases.

\section{Analysis of Fair-NOMA Capacity} \label{sec5}

\subsection{Expected Value of Fair-NOMA Capacity}

The expected value of the Fair-NOMA capacities of MU-1 and MU-2 depend on the power allocation coefficient $a$. In order to determine the bounds of this region, the expected value of each user is derived for the cases of $a=a_\text{sup}$ and $a=a_\text{inf}$. The capacity of each user for OMA is derived to compare with NOMA.
%Since we wish to compare to OMA, the expected value of the capacity is also found for this case. 

Since the channels of two users are i.i.d. random variables,  the joint probability density function is 
\begin{equation}
	f_{|h_1|^2, |h_2|^2}(x_1,x_2) = \frac{2}{\beta^2}e^{-\frac{x_1+x_2}{\beta}}.
\end{equation}
The ergodic capacity of the MU-1's given that MU-1 channel gain is always less than MU-2 channel gain  using OMA is given by
\begin{align}
	\mathbb{E}[ C_1^\text{O} ]  =& \int_0^\infty\int_{x_1}^\infty \frac{2}{\beta^2}e^{-\frac{x_1+x_2}{\beta}}\cdot\frac{1}{2}\log_2(1 + \xi x_1)dx_2 dx_1 \nonumber\\
	=& \frac{e^{\frac{2}{\beta\xi}}}{\ln(4)} E_1\left(\frac{2}{\beta\xi}\right),
\end{align}
where $E_1(x) =\int_x^\infty u^{-1}e^{-u} du$ is the well-known exponential integral. Note that since $C_1^\text{O}=C_1^\text{N}(a_\text{sup})$, their ergodic capacities are also equal. $\mathbb{E}[ C_2^\text{O} ]$ can be derived similarly. 
\begin{equation}
	\mathbb{E}[ C_2^\text{O} ] = \frac{e^{\frac{1}{\beta\xi}}}{\ln(2)}E_1\left(\frac{1}{\beta\xi}\right) - \frac{e^{\frac{2}{\beta\xi}}}{\ln(4)} E_1\left(\frac{2}{\beta\xi}\right)
\end{equation}
Hence, the sum rate capacity of OMA users is 
\begin{equation}
	\mathbb{E}[S_{\text{O}} ] = \frac{e^{\frac{1}{\beta\xi}}}{\ln(2)}E_1\left(\frac{1}{\beta\xi}\right).
\end{equation}
%Therefore, the OMA capacity expected value for MU-2 is 
%\begin{equation}
%	\mathbb{E}[ C_2^\text{O} ] = \frac{e^{\frac{1}{\beta\xi}}}{\ln(2)}E_1\left(\frac{1}{\beta\xi}\right) - \frac{e^{\frac{2}{\beta\xi}}}{\ln(4)} E_1\left(\frac{2}{\beta\xi}\right).
%\end{equation}
Note that in the case that $a=a_\text{inf}$, since $C_2^\text{N}(a_\text{inf}) = C_2^\text{O}$, their ergodic capacities are also equal.

%%%%%%%%%%%%%%%%%%%	NOMA using a_inf

In the case of NOMA using $a= a_\text{inf}$, the capacity of MU-1 is 
\begin{align*}
	&\mathbb{E}\left[C_1^\text{N}(a_\text{inf})\right] = \int_0^\infty\int_0^{x_2}  \frac{2}{\beta^2}e^{-\frac{x_1+x_2}{\beta}}\\
	&\cdot[\log_2(1+\xi x_1)  -\log_2\left(1 + (\sqrt{1+\xi x_2}-1)\frac{x_1}{x_2}\right)]dx_1dx_2.
\end{align*}
This double integral simplifies to the single integral
\begin{align}
	\label{eq:C1ainf} & \mathbb{E}\left[ C_1^\text{N}(a_\text{inf}) \right] =  \frac{3e^{\frac{2}{\beta\xi}}}{\ln(4)}E_1\left(\frac{2}{\beta\xi}\right) \\
	& - \int_0^\infty \frac{2}{\beta\ln(2)}  \exp\left(-\frac{x}{\beta}\left( \frac{\sqrt{1+\xi x}-2}{\sqrt{1+\xi x}-1}\right)\right)\nonumber\\
    & \cdot \left( E_1\left( \frac{x}{\beta(\sqrt{1+\xi x}-1)} \right) - E_1\left(\frac{x\sqrt{1+\xi x}}{\beta(\sqrt{1+\xi x}-1)} \right) \right)  dx \nonumber
\end{align}
which can be calculated by a software such as Matlab.

%%%%%%%%%%%%%%%%%%%%	NOMA using a_sup
Similarly, in the case of NOMA with $a=a_\text{sup}$, the capacity of MU-2 is a given by 
\begin{align*}
	\mathbb{E}[ C_2^\text{N}(a_\text{sup}) ] = & \int_0^\infty\int_{x_1}^\infty  \frac{2}{\beta^2}e^{-\frac{x_1+x_2}{\beta}} \\
	& \cdot \log_2\left(1 + (\sqrt{1+\xi x_1}-1)\frac{x_2}{x_1}\right)dx_2dx_1.
\end{align*}
This double integral simplifies to a single integral of
\begin{align}\label{eq:C2asup}
	&\mathbb{E}[ C_2^\text{N}(a_\text{sup}) ]  = \frac{e^{\frac{2}{\beta\xi}}}{\ln(4)} E_1\left(\frac{2}{\beta\xi}\right) + \int_0^\infty \frac{2}{\beta\ln(2)}  \\ 
	 &\cdot \exp\left(-\frac{x}{\beta}\left(\frac{\sqrt{1+\xi x} -2}{\sqrt{1+\xi x} -1}\right)\right) E_1\left( \frac{x\sqrt{1+\xi x}}{\beta(\sqrt{1+\xi x} - 1)} \right)dx.\nonumber
\end{align}

\subsection{Comparison of Fair-NOMA and OMA}

To observe the gains that are made by using Fair-NOMA, its performance is simulated to confirm the analysis  and compare directly with OMA by varying $\xi$, and setting $\beta=1$. As the analysis demonstrates, there is an increase in the sum capacity when using NOMA compared to OMA.

\begin{figure}
	\centering
	\includegraphics[scale=0.5]{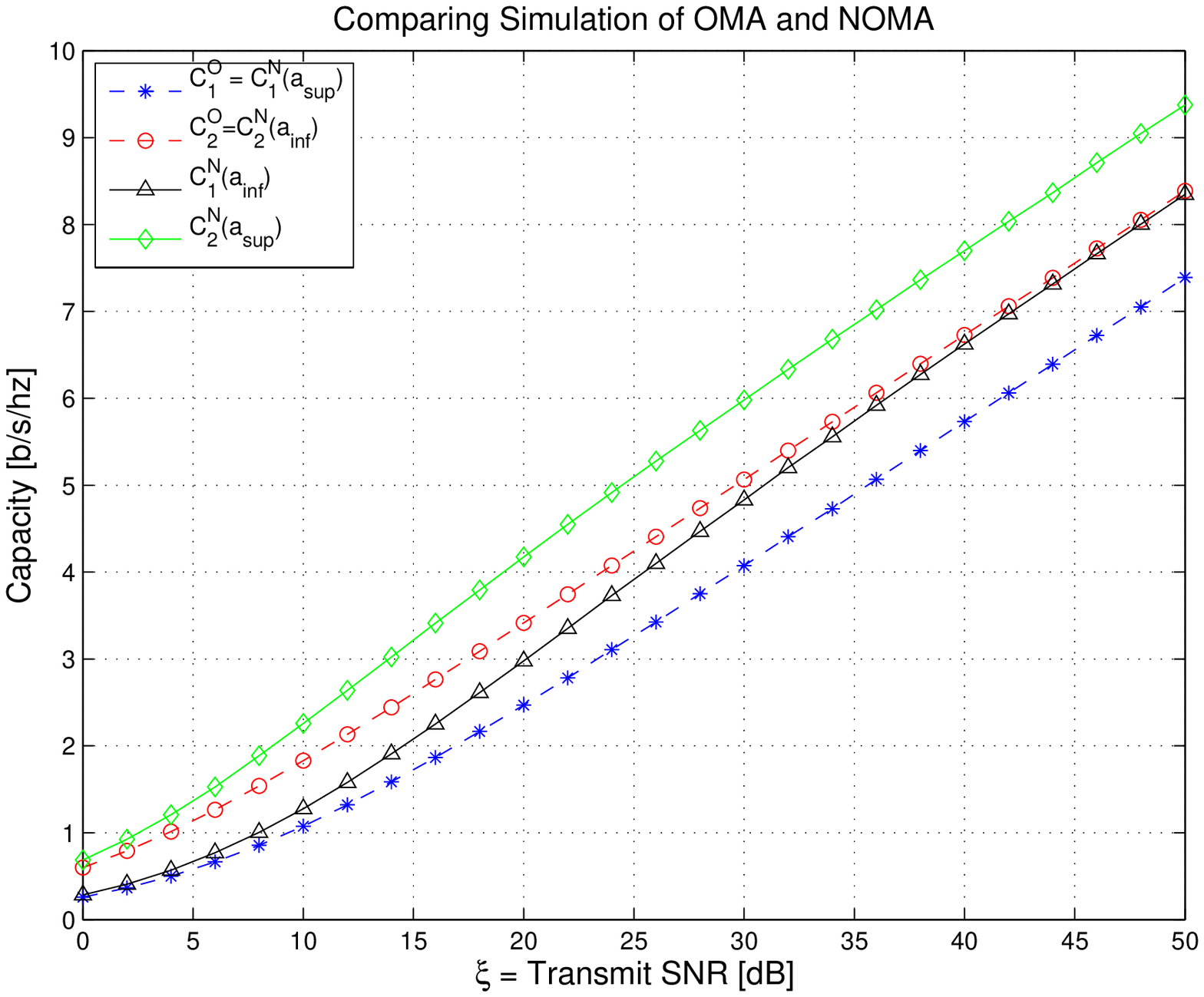}
	\caption{Comparing the capacity of NOMA and OMA}
	\label{fig:nomainc}
\end{figure}

Figure \ref{fig:nomainc} shows NOMA performance for both $a=a_\text{inf}$ and $a=a_\text{sup}$.  Since $C_2^\text{O}=C_2^\text{N}(a_\text{inf})$ and $C_1^\text{O}=C_1^\text{N}(a_\text{sup})$, the plots of $C_1^\text{N}(a_\text{inf})$ and $C_2^\text{N}(a_\text{sup})$ demonstrate the expected capacity increase at the bounds of $\mathcal{A}_\text{FN}$. From the plot, the expected increase in capacity that is made for $a=a_\text{inf}$ is significantly less than the expected increase made for $a=a_\text{sup}$ when the transmit SNR $\xi$ is small. This leads to the conclusion that at lower transmit SNR levels, larger increases in capacity will be observed for the stronger user, and hence the power allocation coefficient should be closer to $a_\text{sup}$. 

However, when the transmit SNR is large ($\xi\geq 30$ dB), the expected increase in capacity for either case of $a$ appears to be the same. Also, the expected capacity of MU-1 seems to be always upper-bounded by the expected capacity of MU-2, but when $a=a_\text{inf}$ the expected capacity of MU-1 asymptotically approaches the expected capacity of MU-2 with increasing $\xi$. This seems to agree with intuition that when the transmit SNR is large, the channel SNR gain will become a less significant factor. What is nice about this result is that the expected capacity increase becomes less dependent on the channel gain of the user, and more dependent on the fact that NOMA is being used to begin with. 
%\begin{figure}[http]
%    \centering
%      \includegraphics[scale=0.35,angle=0]{NOMA_2user_asup_ainf_oma_compare.eps}
%\vspace{-1.0in}
%      \caption{Comparing the capacity of NOMA and OMA}
%    \label{fig:nomainc}
%\end{figure}

\begin{figure}
	\centering
	\includegraphics[scale=0.5]{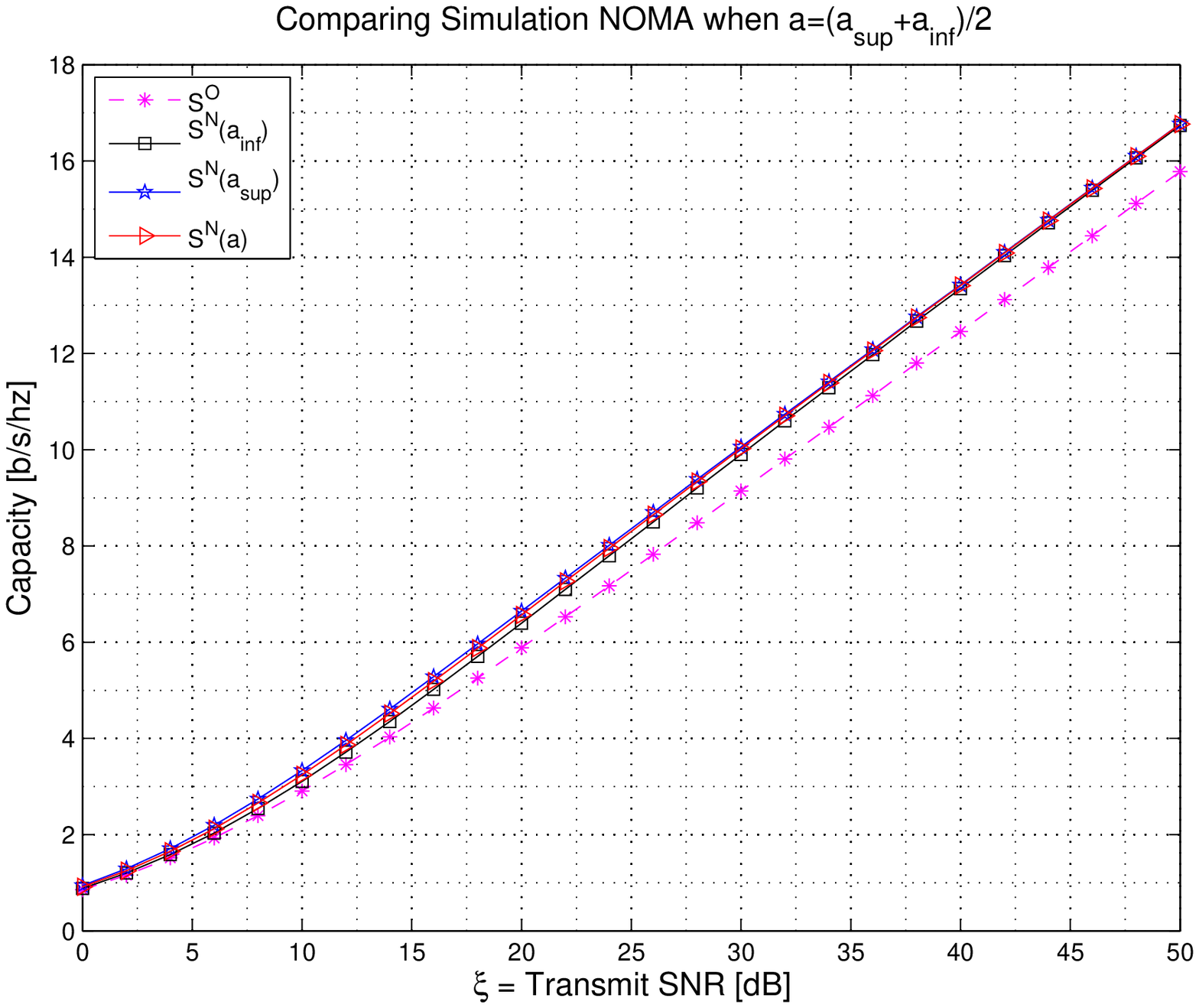}
	\caption{\label{fig:gnoma}Comparing the capacity of NOMA and OMA}
\end{figure}

Given that the expected capacity gain made when the transmit SNR increases is roughly the same for either $a_\text{inf}$ or $a_\text{sup}$, this implies that the expected sum capacity of NOMA should increase by the same amount for all values of $a\in\mathcal{A}_\text{FN}$. This is shown in figure \ref{fig:gnoma}, where three different values of $a\in\mathcal{A}_\text{FN}$ are used to illustrate this fact. The plot also shows that the expected gap between NOMA and OMA in the sum capacity increases steadily as $\xi$ increases, further demonstrating that the effects of NOMA are magnified when $\xi$ is large.
%\begin{figure}[http]
%    \centering
%      \includegraphics[scale=0.35,angle=0]{NOMA_2user_anoma_compare.eps}
%\vspace{-1.0in}
%      \caption{Comparing the sum rate capacity of NOMA and OMA}
%    \label{fig:gnoma}
%\end{figure}

\begin{figure}
	\centering
	\includegraphics[width=\columnwidth]{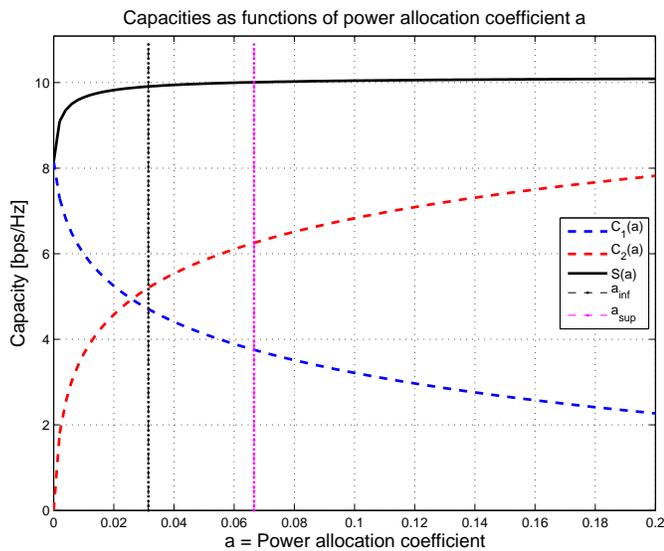}
	\caption{\label{fig:captrade}NOMA capacity tradeoff as a function of $a$; $\xi = 30$ dB}
\end{figure}

Figure \ref{fig:captrade} shows the expected capacity of each MU, and the sum capacity, as functions of the power allocation coefficient $a$ and for $\xi=30$ dB. An interesting observation of this plot is that the largest increase in sum capacity for NOMA occurs in the region  $0<a<a_\text{inf}$ (that is the first vertical line), and then becomes nearly constant once $a\geq a_\text{inf}$ (second vertical line). This is a very promising result, as it means that when transmit SNR is large, there is almost no benefit of using a power allocation coefficient greater than $a_\text{sup}$ in order to attempt to increase the sum capacity, because it is almost near its maximum value. Thus, when it comes to increasing the sum capacity of the system, there is no incentive to allocate more power to the stronger user MU-2, and thus fairness is actually a nearly optimal power operating point. 
%It also implies that approximate knowledge of channel gains are sufficient to maximize the sum capacity.

\section{Conclusion and Future Work}	\label{sec7}

The performance of NOMA when using a fair power allocation coefficient approach, as defined by the power allocation coefficient set $\mathcal{A}_\text{FN}$, was shown to always provide an improvement in system performance. It was shown that the information capacity of each user is always improved when using a power allocation coefficient $a\in\mathcal{A}_\text{FN}$, and that the improvement in capacity is expected to increase as the transmit SNR increases. Moreover, the sum capacity of the system is not improved when the power allocation coefficient favors the stronger user unfairly, and thus fairness in power allocation is desirable. The fact that the sum capacity for NOMA is nearly the same $\forall a\in\mathcal{A}_\text{FN}$ for large values of transmit SNR $\xi$ actually gives flexibility in how to approach maximizing the sum capacity, in the sense that it can be done by focusing on maximizing either the capacity of MU-1 or MU-2. 

%Furthermore, the outage probability was shown to improve for both MUs when using NOMA over OMA. Particularly, the probability of a user selected with weaker channel SNR gain will obtain a significantly lower probability of being in outage, and hence will be allowed to participate with greater probability while still maintaining the same minimum capacity service requirement. Abstractly, this can translate to greater cell coverage in an area, so long as the receivers have SIC capability. The power allocation coefficient can be adjusted in order to strike a balance between optimizing capacity and increasing cell coverage, while always maintaining better performance than would be possible with OMA. 

The next step in this work is  to  analyze the implications of using NOMA fairly, and  extend this concept to more general systems such as MIMO. In a more general system, such as a multi-user MIMO system, the ability to eliminate the need to employ algorithm searches to  find the power allocation that improves capacity becomes necessary, since this can become computationally expensive once the number of antennas in the system grows. The effects of NOMA in systems that employ user pairing approaches should also be investigated, since in these systems the channel SNR gains will no longer be i.i.d., and hence  a different effect in the expected improvements in capacity will be observed.

%%%%%%%%%%%%%%%%%%%%%%%%%%%%%%%%%%%%%%%%%%%%%%%%%%%%%%%
%%%%%%%%%%%%%%%%%%%%%%%%%%%%%%%%%%%%%%%%%%%%%%%%%%%%%%%
% Appendix
%%%%%%%%%%%%%%%%%%%%%%%%%%%%%%%%%%%%%%%%%%%%%%%%%%%%%%%
%%%%%%%%%%%%%%%%%%%%%%%%%%%%%%%%%%%%%%%%%%%%%%%%%%%%%%%

\appendices

%%%%%%%%%%%%%%%%%%%%%%%%%%%%%%%%%%%%%%%%%%%%%%%%%%%%%%%
%%%%%%%%%%%%%%%%%%%%%%%%%%%%%%%%%%%%%%%%%%%%%%%%%%%%%%%
% Appendix A
%%%%%%%%%%%%%%%%%%%%%%%%%%%%%%%%%%%%%%%%%%%%%%%%%%%%%%%
%%%%%%%%%%%%%%%%%%%%%%%%%%%%%%%%%%%%%%%%%%%%%%%%%%%%%%%

%%%%%%%%%%%%%%%%%%%%%%%%%%%%%%%%%%%%%%%%%%%%%%%%%%%%%%%
%%%%%%%%%%%%%%%%%%%%%%%%%%%%%%%%%%%%%%%%%%%%%%%%%%%%%%%
% 					BIBLIOGRAPHY	
%%%%%%%%%%%%%%%%%%%%%%%%%%%%%%%%%%%%%%%%%%%%%%%%%%%%%%%
%%%%%%%%%%%%%%%%%%%%%%%%%%%%%%%%%%%%%%%%%%%%%%%%%%%%%%%

\end{document}